\documentstyle[12pt,aaspp4]{article}

\def\icm{${\rm cm}^{-1}$}
\def\IRAS{{\sl IRAS}}
\def\COBE{{\sl COBE}}
\def\uK{\hbox{$\mu$K}}
\begin{document}

\slugcomment{To appear in Ap.J.}

\title{Detection of Cosmic Microwave Background Anisotropy by the Third Flight of MSAM}
\author{
E.~S.~Cheng\altaffilmark{1},
D.~A.~Cottingham\altaffilmark{2},
D.~J.~Fixsen\altaffilmark{3},
A.~B.~Goldin\altaffilmark{4},
C.~A.~Inman\altaffilmark{4,1},
L.~Knox\altaffilmark{5},
M.~S.~Kowitt\altaffilmark{1,4},
S.~S.~Meyer\altaffilmark{4},
J.~L.~Puchalla\altaffilmark{1},
J.~E.~Ruhl\altaffilmark{4,6},
and~R.~F.~Silverberg\altaffilmark{1}}

\altaffiltext{1}{NASA/Goddard Space Flight Center, Laboratory for Astronomy
and Solar Physics, Code 685.0, Greenbelt, MD 20771}
\altaffiltext{2}{Global Science and Technology, Inc., NASA/GSFC Laboratory
for Astronomy and Solar Physics, Code 685.0, Greenbelt, MD 20771}
\altaffiltext{3}{Applied Research Corporation, NASA/GSFC Laboratory for
Astronomy and Solar Physics, Code 685.0, Greenbelt, MD 20771}
\altaffiltext{4}{University of Chicago, 5640 S. Ellis St., Chicago, IL 
60637} 
\altaffiltext{5}{CITA, University of Toronto, Toronto, Ontario
M5S 3H8, Canada}
\altaffiltext{6}{Department of Physics, University of California,
Santa Barbara, CA 93106}

\begin{abstract}
The third flight of the Medium Scale Anisotropy Measurement
(MSAM1), in June 1995, observed a new strip of
sky, doubling the sky coverage of the original MSAM1 dataset.  
MSAM1 observes with a 0\fdg5 beam size in four bands from 5--20~\icm.
From these four bands we derive measurements of cosmic microwave
background radiation (CMBR) anisotropy and
interstellar dust emission.
Our measurement of
dust emission correlates well with the 100~\micron\ \IRAS\ Sky
Survey Atlas;
from this comparison we determine an effective emissivity spectral index
between 100~\micron\ and 444~\micron\ of $1.46 \pm 0.28$.
Analysis of our measurement of CMBR anisotropy shows that for
Gaussian-shaped correlation functions with $\theta_c = 0\fdg3$,
we place a limit on total rms anisotropy of
$2.2\times10^{-5} < \Delta T/T < 3.9\times10^{-5}$ (90\% confidence
interval, including calibration error).
The band-power limits are $\langle \delta T \rangle \equiv
\langle l (l+1)C_l/2\pi\rangle^{1/2} =
50^{+16}_{-11}\,\uK$ at $l = 160$,
and $\langle \delta T \rangle = 65^{+18}_{-13}\,\uK$ at $l = 270$
($1\,\sigma$ limits, including calibration error).
The corresponding limits with statistical errors only are $\langle
\delta T \rangle = 50^{+13}_{-9}\,\uK$ and $\langle \delta T \rangle
= 65^{+14}_{-10}\,\uK$ respectively.
These measurements are consistent with a standard adiabatic cold dark
matter model;  we discuss constraints on $h$, $n$, and the
redshift of reionization.
\end{abstract}


\section{Introduction}

The Medium Scale Anisotropy Measurement (MSAM) is an experiment to
measure anisotropy in the cosmic microwave background radiation (CMBR) at
angular scales near 0\fdg5.  The first two flights of MSAM1, reported
in \cite{cheng94} (Paper I) and \cite{cheng95} (Paper II), both
observed the same field to demonstrate the repeatability of our
measurements.  The detailed comparison, showing consistency
between these two measurements, was reported in \cite{inman96}.  To
increase sky coverage and consequently increase sensitivity to the CMBR
anisotropy power spectrum, the third flight of MSAM1 (MSAM1-95)
measured a new field using the
same observing method.  This paper reports initial results from
this third flight.  We will report results from the combined dataset
of all three flights in a future paper.

Some recent measurements of CMBR anisotropy
have begun to hint at the shape of the
correlation function near 0\fdg5.  \cite{netterfield96},
from analysis of new measurements from their Saskatoon experiment (SK95),
report a rise from $l \approx 90$ to 240;
\cite{platt96} report that new measurements with Python
III are consistent with a flat power spectrum
from $l \approx 90$ to 180.  Analysis of such measurements
in conjunction with the DMR maps have begun to suggest what area of
cosmological parameter space is viable.  \cite{bond96a} have analyzed
the Saskatoon data along with SP94 and DMR, and find that with a
number of parameters fixed they can put interesting limits on others,
e.g. $H_0$.  In this letter we report similar initial results based on
the MSAM1-95 data.

\section{Instrument and Observations}

A detailed description of the MSAM1 instrument can be found in
\cite{fixsen96a}.  We give a brief overview here.  MSAM is a
balloon-borne 1.4~m off-axis Cassegrain telescope which forms a beam of
width $\sim 0\fdg5$.  A three-position chopping secondary throws the
beam $\pm 0\fdg7$; we run this chopper at 2~Hz.  The MSAM1 detector
system has four spectral channels at 5.7, 9.3, 16.5, and 22.5~\icm,
each with a width of $\sim 1$--2~\icm.  The detectors are sampled 
at 32~Hz: 4
times for each of 4 positions of the secondary mirror, for 
a total of 16 samples per
chopper cycle.  Pointing is determined with a star camera and gyroscope.
The configuration of the gondola for the 1995 flight is unchanged from
the 1994 flight.

The third flight of MSAM1 was launched from the National Scientific
Balloon Facility in Palestine, Texas at 23:54 UT 01 Jun 1995.
CMBR field observations were taken at altitudes of 37.5 -- 40~km.
During the flight, we
observed Mars, Jupiter, and Saturn to calibrate the instrument (these
observations are reported in detail in \cite{goldin96}).  To measure
anisotropy in the CMBR, we observed a strip of sky at $\delta \sim
80\fdg5$ from $\alpha \sim 14\fh2$ to 19\fh5
over the period 04:20 to 09:22 UT.  
We terminated the observation at that time because we were entering
a region of increased emission from Galactic dust.
This field parallels the
MSAM1-92/MSAM1-94 field at a distance of 1\fdg5 in declination.

The observing method for our CMBR scans is the same as that described
in papers I and II, adjusted for the faster sky motion at the lower
declination.  We begin with the telescope pointed at a spot at $\delta
= 80\fdg5$, 25\arcmin\ east of the meridian.  We scan in azimuth a
distance of $\pm 47\arcmin$ on the sky with a
period of 60~s and track sky rotation until the center of the scan is
25\arcmin\ west of the meridian.  This takes about 20~min.  We then
pause briefly for a star camera picture, move 50\arcmin\ to the right
(so we are again 25\arcmin\ east of the meridian), and start a new scan.
During the flight we completed 15 scans (integration time on the first
scan was only 14~min).

\section{Data Analysis}

The reduction of the data consists of the following steps.
The detector data are contaminated by spikes caused by
cosmic rays striking the detectors; we remove these spikes.  The data
are calibrated by our observation of planets, and are
analyzed to provide measurements of brightness in our four spectral
channels as a function of sky position.  These are then fit to a
spectral model to produce measurements of CMBR anisotropy and dust
optical depth.  These analyses and their results are described in the
following sections.

\subsection{Detector Data Reduction}

The observed signal is noise plus a convolution of the radiation
incident on the bolometer with the transfer function of the
bolometers, electronics and sampling.  To reconstruct the incident
signal optimally (with minimum variance) from that observed we apply a
Wiener filter to the data.
The detector data contain spikes,
generally consistent with cosmic rays
striking the bolometers.  
In the unfiltered time stream, a cosmic ray hit contaminates $\sim 30$
time samples.  The Wiener filter compresses this mostly into one
sample with some ringing present in the adjacent samples --- all of
which are cut.  The extent of the ringing and therefore the breadth of
the cut depends on the glitch amplitude; a typical breadth of the cut
is 10 samples.
In addition, an interval of 12~s around each telemetry
dropout is removed.  A total of 11\% of the data is deleted in this
step: 10\% for spikes, and 1\% for telemetry dropouts.
Next, certain periods of data have excess noise, correlated with noise in 
the data from a
cryogenic temperature sensor.
This suggests an intermittent electrical problem as the common origin.
An additional 13\% of the data are cut based on the noise level in
this sensor.  Finally, we delete any
chopper cycle in which any points are cut,
thus eliminating an additional 6\% of the data,
for a total of 30\% of the data deleted.  
The fractions of data deleted given above
are relative to the total 5.0~h period of
CMBR observations.

We perform two demodulations of the detector signals.  
Let $T_L$, $T_C$, and $T_R$ be the sky temperature at the left,
center, and right position of the chopper respectively.
The single difference demodulation is $T_R - T_L$,
making an antisymmetric beam pattern.  The double difference is
$T_C - (T_L + T_R) / 2$,
making a symmetric beam-pattern.  We use our
scan over Jupiter to choose demodulations optimized for signal to
noise ratio.

We estimate the instrument noise by measuring the variance in 100~s
segments of the demodulated data after removing a slow
drift.  
For the double difference demodulation,
the achieved sensitivity in each channel is 180, 160, 110, and
$160\,\uK\,{\rm s}^{1/2}$ Rayleigh-Jeans; for the single difference
it is 20\% to 50\% larger.  For channels 1 and 2 this corresponds to
360 and $910\,\uK\,{\rm s}^{1/2}$ CMBR.
The offset in the demodulated data in the various channels and
demodulations ranges from 1 to 5~mK; the offset drift over the whole
flight is small
compared to this.

We divide the data into 0\fdg12 bins in sky position
along RA and Dec, and 10\arcdeg\ bins
in angular orientation of the chopper direction on the sky (roll).  We
dedrift and bin the data by fitting the data for each channel to a
model consisting of a brightness in each bin, plus a slowly varying
function of time (a cubic spline with knots spaced at 5~min
intervals).  This is done separately for each channel and demodulation.
The reduced $\chi^2$ of these fits are 0.99--1.17.

The data are calibrated by our observations of Jupiter, using the values
of the brightness temperature of Jupiter reported in \cite{goldin96}
(of the two models presented in that paper, we use
the temperatures based on the ``Rudy model,'' \cite{rudy87b}).
The error in the
calibration is estimated to be 5\%,
dominated by the uncertainty in the Jupiter temperature.

\subsection{Spectral Decomposition}

We derive measurements of CMBR anisotropy and ISM dust emission by
fitting the
four channels to a
two-component spectral model.  The first component is cosmic microwave
background radiation with $T = 2.728$~K (\cite{fixsen96b}), with
the free parameter being the anisotropy $\Delta T$; and the second
component is emission from dust in the interstellar medium, with the
free parameter being the optical depth.  The dust is assumed to have a
temperature of 20~K and an emissivity spectral index of 1.5; we find
that varying the spectral index from 1.3 to 2.0 has essentially no
effect on the CMBR results, but does affect the dust results as noted
below.
The $\chi^2/{\rm DOF}$ of this fit is
505/548 for the single difference demodulation, and 533/548 for the
double difference, both entirely consistent with a $\chi^2$
distribution.

\begin{figure}[tp]\begin{center}
\plotone{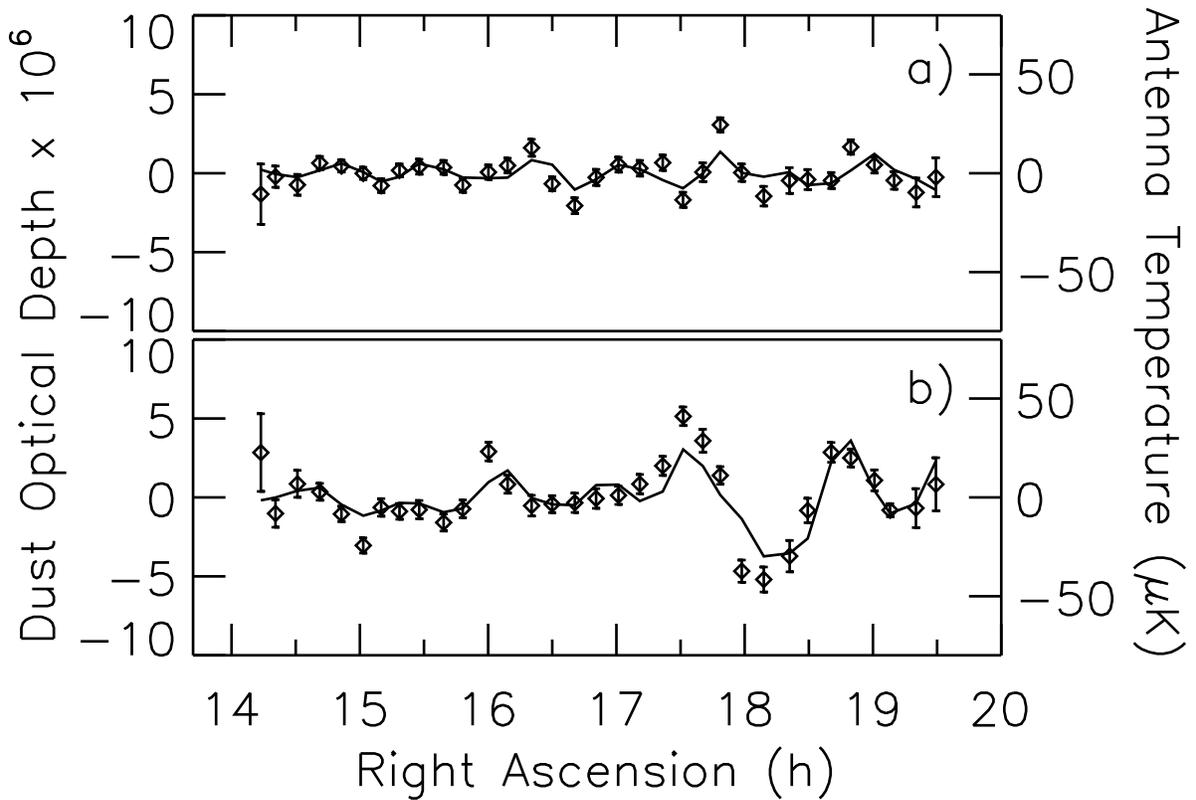}\caption{ Optical depth of interstellar dust at 22.5~\icm.
The right-hand scale is antenna temperature at 22.5~\icm.
The points with error bars are our measurements.  The curve is the
ISSA 100~\micron\ data convolved with our beam pattern, with amplitude
fitted to our measurements.  a) Double difference demodulation; b)
single difference.
\label{f_dust} }
\end{center}\end{figure}

The resulting measurement of interstellar dust optical depth is
plotted in Fig.~\ref{f_dust}.
We have fit these measurements to the \IRAS\ Sky Survey Atlas (ISSA)
at 100~\micron\ (\cite{wheelock93}) convolved with our beam patterns.
In this fit we assume a dust temperature of 20~K, and find the
best-fit optical depth ratio between 100~\micron\ and 444~\micron.
Fig.~\ref{f_dust} also shows this ISSA model.  The $\chi^2$ of this
fit is 449/273 for the single difference demodulation, and 356/273 for
the double difference.  There is clear correlation between this
measurement and the ISSA, but the improbable values of $\chi^2$
indicate that they are not in complete agreement.  From the results of
this fit we can assign an effective dust emissivity spectral index
between 100~\micron\ and 444~\micron\ of $1.46\pm 0.28$; the error bar
includes the effect of varying the assumed index in the spectral
decomposition from 1.3 to 2.0.
This measurement of index is
consistent with our previous measurements (Papers I and II),
and with measurements by the \COBE\ FIRAS and DIRBE instruments over
the entire sky (\cite{reach95}).

\begin{figure}[tp]\begin{center}
\plotone{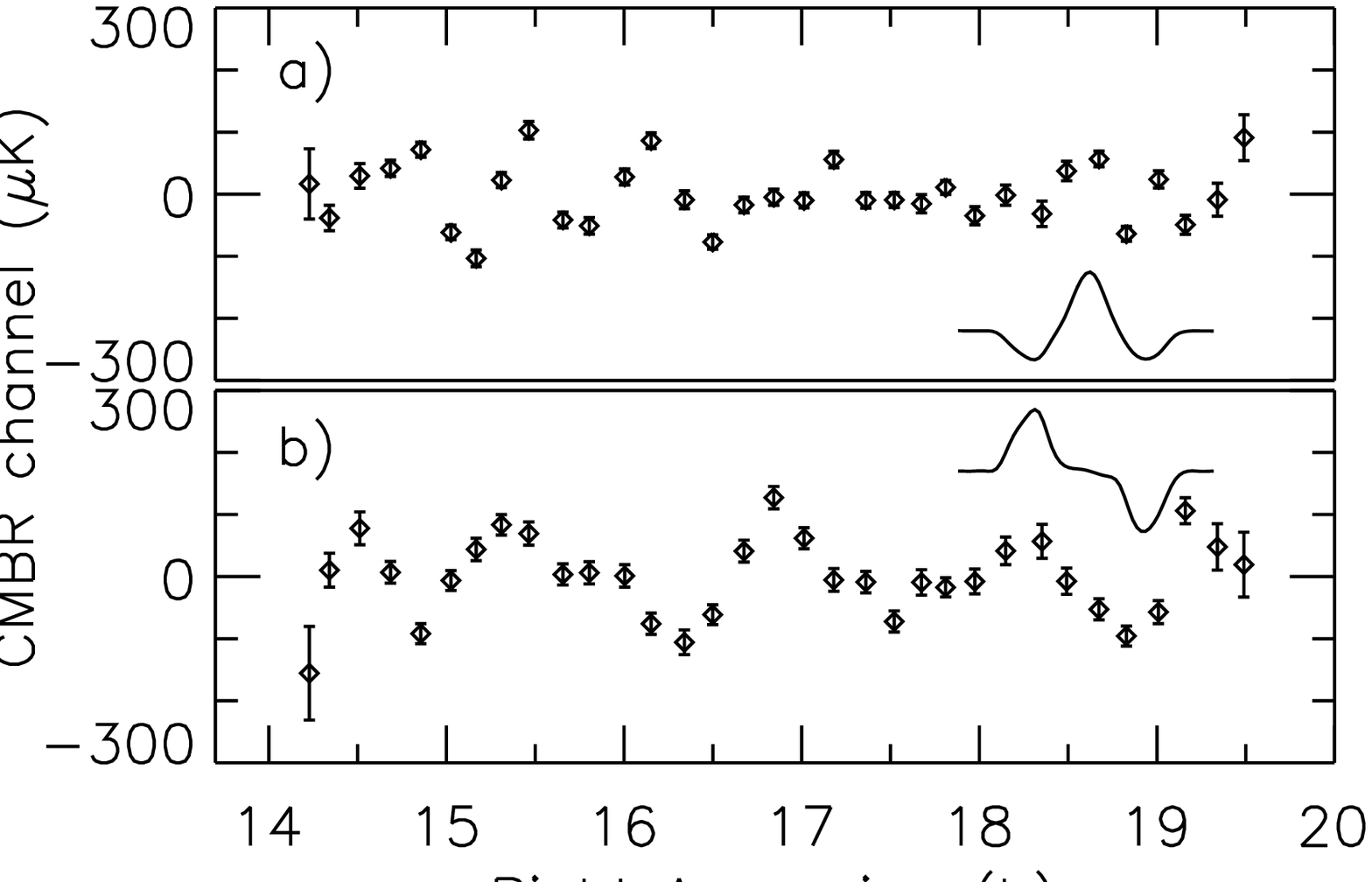}\caption{Our measurements of CMBR anisotropy.  For reference,
the beam pattern is superposed.  a) Double difference demodulation; b)
single difference.
\label{f_cmbr} }
\end{center}\end{figure}

The CMBR component is plotted in Fig.~\ref{f_cmbr}.
For reference, the antenna pattern is also shown.
The anisotropy is clearly visible.
This dataset is considered further in the next subsection.

The procedure for producing figures \ref{f_dust} and \ref{f_cmbr}
needs to be briefly explained.  The dataset that results from our
analysis is difficult to represent in a plot for three reasons: there
are a large number of data points (275), it has a significantly
non-diagonal covariance matrix, and it is a function of declination
and roll as well as of right ascension.  To make the figures, we do the
following.  First, we project out the eigenmodes of the covariance
matrix with the largest eigenvalues ($\sim 10$), from both the data and the
covariance.  This results in a nearly diagonal covariance, from which
we estimate the error bars.  We then rebin the data, ignoring
declination and roll, and using coarser right ascension bins.  These
steps are performed only for making a representative and comprehensible visual
presentation of the data; all analyses are done with the full
dataset and covariance matrix.

\subsection{CMBR Anisotropy}

To assess the implications of these observations for cosmological
parameters, we have used the same method as in Papers I and II.
We choose a correlation function to test, and use the likelihood ratio
statistic to find a 90\% confidence interval on an overall multiplier
of the correlation function.  This confidence interval is bounded by
the 95\% confidence level upper and lower bounds.
As in papers I and II, we have applied
this test to Gaussian-shaped correlation functions of various
correlation lengths.  We have also tested a number of CDM models.
In Papers I and II, we only presented results based on each
demodulation separately; here we also present results based on both
taken together.

\begin{figure}[tp]\begin{center}
\plotone{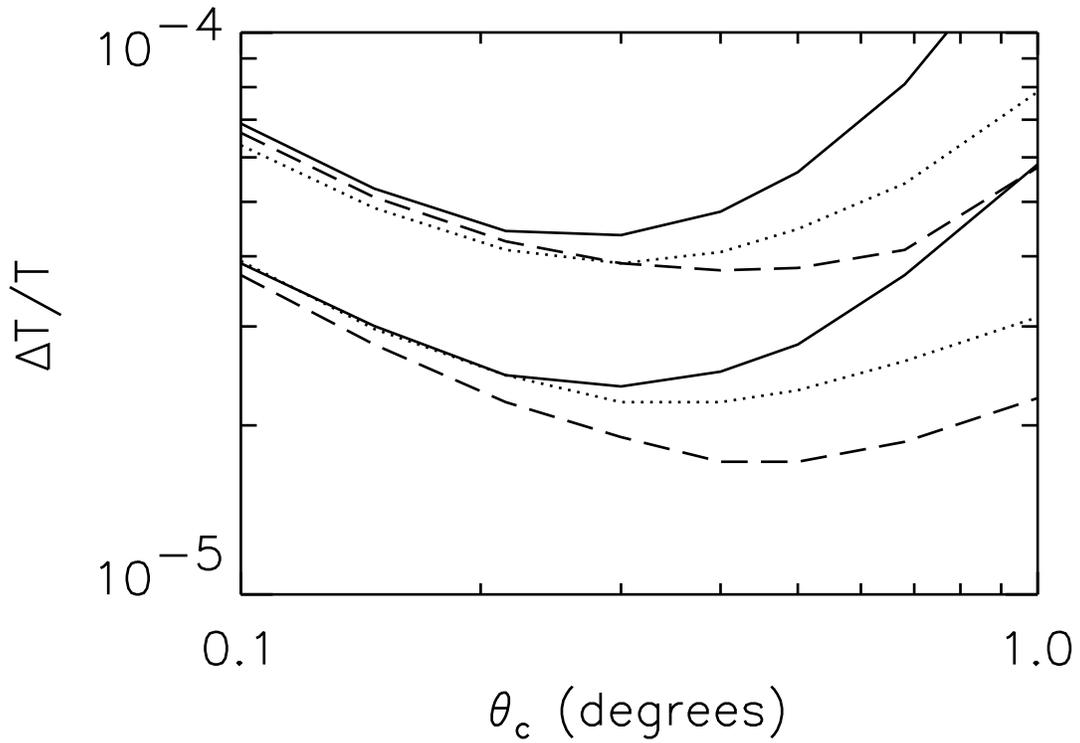}\caption{Limit on total rms anisotropy for Gaussian-shaped
correlation functions.  Each limit is a 95\% confidence level; an
upper and lower bound together limit a 90\% confidence interval.
Shown are upper and lower limits for the double
difference demodulation (solid lines); upper and lower limits
for the single difference demodulation (dashed); and upper and lower
limits for both demodulations (dotted).
The calibration error is not included in these limits.
\label{f_deltat} }
\end{center}\end{figure}

The results from the Gaussian-shaped correlation functions,
$C(\theta) \propto \exp[-(\theta/\theta_c)^2/2]$,
are shown
in Fig.~\ref{f_deltat}.
This figure shows limits for each demodulation, and for both demodulations
taken together.  The horizontal axis is the correlation
length $\theta_c$ of the correlation function, and the vertical axis
is total rms anisotropy, $[C(0)]^{1/2}$.  Table~\ref{t_deltat} gives
results for selected correlation lengths.
Both the figure and the table show 90\% confidence intervals.

We have also tested various adiabatic cold dark matter
models using correlation functions
computed with CMBFAST (\cite{seljak96}).  Starting from a nominal set
of parameter values, we vary one parameter at a time, and
find the interval of that parameter which produces correlation
functions consistent with these measurements.
The nominal set of parameters is $h = 0.5$, $\Omega = 1$,
$\Omega_b h^2 = 0.0125$, $\Lambda = 0$, $Y_{\rm He} = 0.24$, $n = 1$,
and no early reionization, using only scalar perturbations.
CMBFAST produces correlation functions which are
normalized to DMR via the fitting function of \cite{bunn97}; this
normalization has an uncertainty of 7\%.  
We test correlation functions
in the same way as above, with an overall multiplier as the free
parameter.  If the 90\% confidence interval on this parameter,
expanded by the combined calibration uncertainty of this experiment
and the DMR normalization,
does not include the value 1, then we
reject that correlation function.  To be conservative, we combine the
normalization and calibration errors by adding, yielding 12\%.
In varying $h$, holding the other parameters fixed, we
find that these data are consistent with $h < 0.78$.
In varying the spectral index $n$ of the initial perturbations,
we find that it is constrained to $ 0.8 < n < 1.35$.
To examine early reionization, we adopt a model that the ionization
fraction goes suddenly from 0 to 1 at some redshift $z$; we find that
$z < 56$.  This corresponds to an optical depth $\tau$ to the last
scattering surface of $\tau < 0.37$.

We compute the band power estimator of Bond (\cite{bond95}) by hypothesizing
a correlation function $C_l = 6 C_2 / l (l + 1)$ and using the above
procedure to place limits on $C_2$; then the band power estimator $\langle
\delta T \rangle \equiv
\langle l (l+1)C_l/2\pi\rangle^{1/2} = \sqrt{3/\pi} C_2^{1/2}$.
To produce $1\,\sigma$ bounds,
we compute 84\% confidence level upper and lower bounds; the central
value is the median (50\% confidence level bound).  For the single
difference demodulation, we find that $\langle \delta T \rangle =
50^{+16}_{-11}\,\uK$, at a mean $l = 160$; and for the double
difference demodulation, $\langle \delta T \rangle =
65^{+18}_{-13}\,\uK$, at a mean $l = 270$ (both 1~$\sigma$
limits, including calibration error). 
We include calibration error in our limits by adding it to the
statistical error (not in quadrature).  The $1\,\sigma$ limits with
statistical errors only are $50^{+13}_{-9}$ and $65^{+14}_{-10}\,\uK$
respectively.

\section{Conclusions}

The MSAM1-95 data show a highly significant detection of CMBR anisotropy,
consistent in amplitude with the MSAM1-92/MSAM1-94 measurements
(Papers I and II).
With our calibration error taken into
account, the limits in Table~\ref{t_deltat} for both demodulation and
the full dataset are, for Gaussian-shaped correlations functions
with $\theta_c = 0\fdg3$, $2.2 \times
10^{-5} < \Delta T / T < 3.9 \times 10^{-5}$ (90\% confidence
interval).

In Paper II, we reserved judgment about the repeatability of our
single difference measurement, but \cite{inman96} have shown that,
for both demodulations, observations of the same field in 1992 and
1994 agree.  This is a strong argument in favor of the reliability of
MSAM1 data; therefore we now recommend without reservation the use of
results based on both demodulations.

The band-power estimates for the single and double difference quoted
above suggest a rise in the power spectrum from $l = 160$ to $270$.
However, the band-power estimate we quote for the double difference
($l = 270$) is
$1.78\,\sigma$ larger than the MSAM1-94 measurement (Paper II); that
measurement, converted to these units, is $\langle \delta T \rangle =
40^{+12}_{-10}\,\uK$.  The probability of a difference of this
magnitude or greater is 7.5\%, not small enough to call this an
inconsistency.  In this light we might ask if current
measurements of band-power in this range of angular scale lend any
weight to the hypothesis that the power spectrum is rising.
Our present measurement of band-power at $l = 160$ is
in agreement with the measurement of MSAM1-94, $\langle \delta T
\rangle = 35^{+14}_{-10}\,\uK$, and with the measurement by
Python, $\langle \delta T \rangle = 66^{+17}_{-16}\,\uK$ at $l = 170$
(\cite{platt96}).  Our measurement at $l = 270$ is in agreement with
that by SK95, $\langle \delta T \rangle = 85^{+23}_{-17}\,\uK$ at $l
= 240$ (\cite{netterfield96}, error bar adjusted to include
calibration uncertainty); but as noted above, our MSAM1-94
measurement, while statistically consistent with these numbers,
is rather smaller.  These three experiments, taken together, suggest
that there is a rise in the power spectrum from $l \sim 160$ to $l
\sim 270$, but the statistical significance is modest.  If we choose to
ignore MSAM1-94, the suggestion is stronger, but the significance is
still not compelling.

Our measurements of CMBR and dust will be made available on our FTP
server, in {\tt ftp://cobi.gsfc.nasa.gov/pub/data/msam-jun95}.

\acknowledgements

This work would not be possible without the excellent support we
receive from the staff of the National Scientific Balloon Facility.
We thank G. Hinshaw and G. Wilson for illuminating discussions.
Financial support was provided by the NASA Office of Space Science,
under the
theme ``Structure and Evolution of the Universe.''

\clearpage
\bibliographystyle{aas}
\bibliography{cmbr}

\clearpage

\clearpage

\begin{deluxetable}{rcrr}
\tablecolumns{4}
\tablecaption{Upper and lower bounds on total rms CMBR anisotropy ($\protect\sqrt{C_0}$) 
\label{t_deltat} }
\tablehead{
\colhead{} &	\colhead{} &	\colhead{Upper} & \colhead{Lower} \\
\colhead{$\theta_c$} &
		\colhead{Demodulation} &
				\colhead{Bound} & \colhead{Bound}	\\
\colhead{} &	\colhead{} &	\colhead{(\uK)} & \colhead{(\uK)}	}

\startdata
0\fdg5 &	Single &	104 &		47 \nl
0\fdg3 &	Double &	118 &		63 \nl
0\fdg5 &	Both &		122 &		64 \nl
0\fdg3 &	Both &		107 &		61 \nl
\enddata

\tablecomments{The limits in this table do not include calibration
uncertainty.}
\end{deluxetable}

\end{document}